\documentclass[twocolumn,showpacs,reprint]{revtex4-1}
\usepackage{graphicx,epsf}
\usepackage{color}
\usepackage{bm}
\usepackage{dcolumn}
\usepackage{amsfonts}
\usepackage{mathrsfs}

\usepackage{subfig}

\begin{document}

\title{On beat-driven and spontaneous excitations of zonal flows by drift waves}

\author{ Liu Chen$^{1,2}$, Zhiyong Qiu$^{1,2}$, and Fulvio Zonca$^{2, 1}$}

\affiliation{$^1$Institute for    Fusion Theory and Simulation, School  of Physics, Zhejiang University, Hangzhou, P.R.C\\
$^2$ Center for Nonlinear Plasma Science and   C. R. ENEA   Frascati, C.P. 65, 00044 Frascati,  Italy}

\begin{abstract}
Using the slab plasma as a paradigm model, we have derived analytically equations for the nonlinear generation of zero-frequency zonal flows by electron drift waves including, on the same footing, both the beat-driven and spontaneous excitations. It is found that the  beat-driven zonal flow tends to reduce the frequency  mismatch between the electron drift waves and, thereby, contributes to a significant $O(1)$ enhancement of  the modulational instability drive  and lowering its threshold. Implications to tokamaks plasmas as well as drift-wave soliton formation  are also discussed. 
\end{abstract}


\maketitle


Zero-frequency zonal flow  (ZF) \cite{AHasegawaPoF1979,MRosenbluthPRL1998,ZLinScience1998} or, more generally, zonal state (ZS) \cite{FZoncaJPCS2021,MFalessiNJP2023} are known to play crucial roles in regulating low-frequency electromagnetic fluctuations; such as drift waves (DWs)  and shear Alfv\'en waves (SAWs), and the associated plasma transport  \cite{ZLinScience1998,LChenPoP2000,LChenRMP2016}. It is, thus, important to understand the generating mechanisms of ZF  by electromagnetic turbulences. Theoretically, it has been known that ZF can be spontaneously excited via modulational instabilities, due to nonlinear coupling between DW/SAW sidebands \cite{LChenPoP2000,PDiamondPPCF2005,LChenPRL2012,ZQiuNF2017}. Simulations, particularly, on Alfv\'en eigenmodes, however, have also observed, especially during the initial linear stage, generation of zonal fields via self beating of the primary electromagnetic waves; i.e., a  thresholdless  forced-driven mechanism \cite{YTodoNF2010,GDongPoP2019,ABiancalaniJPP2020}.  Hereafter, we shall term it as beat-driven in order to describe more precisely its generation mechanism.   It is, thus, desirable to develop a theoretical framework where both the beat-driven and spontaneous modulational instability generation mechanisms can be treated on the same footing, and to explore its implications and consequences. This is the primary motivation for the present work.


In order to simplify the analysis and elucidate more clearly the underlying physics mechanisms, we consider here, as a paradigm model, nonlinear interactions  between electrostatic electron drift waves (eDWs) and zonal flow  in a slab plasma with $\mathbf{B}=B_0\mathbf{\hat{z}}$ and nonuniformities in $\hat{\mathbf x}$. Meanwhile,  the background plasma is taken to have Maxwellian distributions for both the electrons and the singly-charged ions. 

The governing equations are then given by the nonlinear gyrokinetic  equation \cite{EFriemanPoF1982} and the quasi-neutrality condition. That is, for $j=e, i$ and in terms  of the Fourier-mode representation, 
\begin{eqnarray}
\delta f_{kj}=-\left(\frac{e}{T}\right)_jF_{Mj}\delta\phi_k + e^{-i\bm{\rho}\cdot\mathbf{k}_{\perp}} \delta g_{kj}.\label{eq:deltaf}
\end{eqnarray}
Here, $\delta f_k$ and $\delta \phi_k$ are, respectively, the perturbed distribution function and scalar potential, $\bm{\rho}_j\equiv \mathbf{b}\times\mathbf{v}/\Omega_{cj}$, $\mathbf{b}=\mathbf{B}_0/B_0=\hat{\mathbf{z}}$, and $\Omega_{cj}=(eB_0/mc)_j$.  The nonadiabatic response $\delta g_{kj}$, meanwhile, consists of linear and nonlinear components; i.e., $\delta g_{kj}=\delta g^{(1)}_{kj}+\delta g^{(2)}_{kj}$, given by
\begin{eqnarray}
i(k_{\parallel}v_{\parallel}-\omega_k)\delta g^{(1)}_{kj}=-i(\omega-\omega_{*j})_k(e/T)_j F_{Mj}J_k\delta\phi_k,\label{eq:gk_linear}
\end{eqnarray}
and
\begin{eqnarray}
i(k_{\parallel}v_{\parallel}-\omega_k)\delta g^{(2)}_{kj}=-(c/B_0)\Lambda^{k'}_{k''}\left(J_{k'}\delta\phi_{k'}\delta g_{k''}\right)_j.\label{eq:gk_NL}
\end{eqnarray}
Here, we have taken $\mathbf{k}=k_y\hat{\mathbf{y}} +k_x\hat{\mathbf{x}}+ k_{\parallel}\mathbf{b}$, $k_x=-i\partial_x$, $\omega_{*j}F_{Mj}=v^2_{tj}(\mathbf{k}\times\mathbf{b}/\Omega_{cj})\cdot\nabla F_{Mj}$, $v_{tj}=\sqrt{T_j/m_j}$, $J_k=J_0(k_{\perp}\rho)$ with $J_0$ being the zeroth-order Bessel-function, and $\Lambda^{k'}_{k''}=\mathbf{b}\cdot\mathbf{k''}\times\mathbf{k'}$ satisfying $\mathbf{k}=\mathbf{k'}+\mathbf{k''}$. The quasi-neutrality condition,  $\sum_{j=e,i}\langle e\delta f\rangle_j\simeq 0$, then becomes
\begin{eqnarray}
\frac{N_0e^2}{T_e}\left(1+\frac{T_e}{T_i}\right)\delta\phi_k=\sum_{j=e,i} \langle e J_k\delta g_k\rangle_j,\label{eq:qn}
\end{eqnarray}
where $N_0$ is the equilibrium particle density, and  $\langle \cdots\rangle_j$ denotes integration over the velocity space. 

We now let $\delta\phi_k$ to be composed of a coherent drift-wave and zonal components; i.e., 
\begin{eqnarray}
\delta\phi_k=\delta\phi_d+\delta\phi_Z,\label{eq:nonlinear}
\end{eqnarray}
where,
\begin{eqnarray}
\delta\phi_d=\phi_d(x,t)\exp(i(k_y y+k_{\parallel} z -\omega_{dr} t)) + c.c.,
\end{eqnarray}
and
\begin{eqnarray}
\delta\phi_Z=\phi_Z(x,t)+ c.c..
\end{eqnarray}

From Eq. (\ref{eq:deltaf}),  we then have,  for the drift wave and noting $|k_{\parallel}v_{te}|\gg|\omega_k|\gg|k_{\parallel}v_{ti}|$ and $|k_{\perp}\rho_e|\ll1$, the following linear responses;
\begin{eqnarray}
\delta g^{(1)}_{ki}\simeq \left(1-\frac{\omega_{*i}}{\omega}\right)_k \frac{e}{T_i} J_kF_{Mi}\delta\phi_k,\label{eq:dw_linear_i}
\end{eqnarray}
and 
\begin{eqnarray}
\delta g^{(1)}_{ke}\simeq O(\omega/k_{\parallel}v_{te})\ll1.\label{eq:dw_linear_e}
\end{eqnarray}
Note, in Eq. (\ref{eq:dw_linear_i}), $\omega_k=\omega_{kr}+i \partial_t$. Meanwhile, for the zonal component, we have
\begin{eqnarray}
\delta g^{(1)}_{Zi}=\frac{e}{T_i} F_{Mi}J_Z\delta\phi_Z,\label{eq:zf_linear_i}
\end{eqnarray}
and
\begin{eqnarray}
\delta g^{(1)}_{Ze}=-\frac{e}{T_e} F_{Me}\delta\phi_Z.
\end{eqnarray}

Next, we consider the nonlinear generation of the zonal fluctuations by the drift waves. From Eqs. (\ref{eq:gk_NL}) and (\ref{eq:dw_linear_e}), we have
\begin{eqnarray}
\delta g^{(2)}_{Ze}\simeq 0;
\end{eqnarray}
and, thus, electrons make negligible nonlinear contributions. As to ions, we have
\begin{eqnarray}
\frac{\partial}{\partial t}\delta g^{(2)}_{Zi}=-\frac{c}{B_0}\Lambda^{k'}_{k''}\left(J_{k'}\delta\phi_{k'}\delta g_{k''}\right)_i. \label{eq:zf_nonlinear_i}
\end{eqnarray}
Here, $k'_y+k''_y=0$ and $k'_{\parallel}+k''_{\parallel}=0$. The nonlinear coupling terms in Eq.  (\ref{eq:zf_nonlinear_i}), meanwhile, consist of two types of nonlinear interactions.  The first one is the beat-driven process due to the ponderomotive force produced by the self beating of the eDWs. In this case, one has $|k'_x|=|k''_x|$ or, equivalently, $|k'_{\perp}|^2=|k''_{\perp}|^2$. The second one is spontaneous excitation produced by the nonlinear interactions between the radial sidebands. Thus, in this case, $|k'_x|\neq |k''_x|$ or, equivalently, $|k'_{\perp}|^2\neq |k''_{\perp}|^2$. Denoting, correspondingly, 
\begin{eqnarray}
\delta g^{(2)}_{Zi}=\delta g^{(2)}_{Zi,A}+\delta g^{(2)}_{Zi,B},
\end{eqnarray}
and noting Eqs.   (\ref{eq:dw_linear_i}) and (\ref{eq:zf_nonlinear_i}), we then obtain the beat-driven response as
\begin{eqnarray}
\delta g^{(2)}_{Zi,A}=\frac{c}{B_0} k_y J^2_k \left(\frac{\omega_{*i}}{\omega^2_r}\right)_d\frac{e}{T_i} F_{Mi} \frac{\partial}{\partial x}\left|\phi_d\right|^2. \label{eq:zf_nl_i_a}
\end{eqnarray}
In deriving Eq. (\ref{eq:zf_nl_i_a}), we have noted $\Lambda^{k'}_{k''}\delta\phi_{k'}\delta\phi_{k''}=-ik_y\partial_x(\delta\phi_{k'}\delta\phi_{k''})$ and $\delta\phi_{k'}\omega_{k''}\delta\phi_{k''} + \delta\phi_{k''}\omega_{k'}\delta\phi_{k'} = i\partial_t(\delta\phi_{k'}\delta\phi_{k''})$. Meanwhile, for $\delta g^{(2)}_{Zi,B}$, we can express it as \cite{LChenRMP2016}
\begin{eqnarray}
\frac{\partial}{\partial t} J_Z\delta g^{(2)}_{Zi,B}&=&i\frac{c}{B_0}k_y\frac{\partial}{\partial x}\left[(J_ZJ_{k'}-J_{k''}+J_{k''})_i\delta g_{k''i}\delta\phi_{k'}\right.\nonumber\\
&&\left. - (J_ZJ_{k''}-J_{k'}+J_{k'})_i\delta g_{k'i}\delta\phi_{k''}\right].\label{eq:zf_nl_i_b}
\end{eqnarray}
The zonal component of the quasi-neutrality condition, Eq. (\ref{eq:qn}), along with Eqs. (\ref{eq:zf_linear_i}) - (\ref{eq:zf_nl_i_b}) then yields
\begin{eqnarray}
\frac{N_0e^2}{T_i}\left(1-\Gamma_Z\right)\delta\phi_Z=\left\langle e J_Z\left(\delta g^{(2)}_{Zi,A} + \delta g^{(2)}_{Zi,B}\right)\right\rangle.\label{eq:qn_zf}
\end{eqnarray}
Here, $\Gamma_k=I_0(b_k) \exp(-b_k)$ with $I_0$ being the modified Bessel function and $b_k=k^2_{\perp}\rho^2_i=k^2_{\perp}v^2_{ti}/\Omega^2_{ci}$. Letting $\phi_{Zb}$ and $\phi_{Zs}$ denoting, respectively, the beat-driven and spontaneously excited zonal potentials, we then have, taking the $b_k\ll1$ limit, $\phi_Z=\phi_{Zb}+\phi_{Zs}$, where, noting Eqs. (\ref{eq:zf_nl_i_a}) and (\ref{eq:zf_nl_i_b}),
\begin{eqnarray}
\frac{\partial^2}{\partial x^2} \phi_{Zb}=-\frac{c}{B_0}k_y\frac{\omega_{*in}}{\omega^2_{dr}\rho^2_{i}} \frac{\partial}{\partial x}\left|\phi_d\right|^2, \label{eq:zf_fd}
\end{eqnarray}
and
\begin{eqnarray}
&&\frac{\partial}{\partial t}\left(\frac{\partial^2}{\partial x^2}\phi_{Zs}\right)=-\frac{T_i}{N_0e^2\rho^2_{i}}\frac{\partial}{\partial t} \left\langle eJ_Z\delta g^{(2)}_{Zi,B} \right\rangle\nonumber\\
&\simeq& i\frac{c}{B_0} k_y  \alpha_i \frac{\partial^2}{\partial x^2}\left(\phi_d\frac{\partial}{\partial x}\phi^*_d - \phi^*_d\frac{\partial}{\partial x}\phi_d\right).\label{eq:zf_mi}
\end{eqnarray}
In deriving Eq. (\ref{eq:zf_mi}), we have employed the quasi-neutrality condition for $\langle e J_k\delta g_k\rangle_i$ as well as approximated $J_k\simeq 1-b_k/4$ and $\delta g_{ki}\simeq \delta g^{(1)}_{ki}$ in Eq. (\ref{eq:zf_nl_i_b}).  Also,  $\omega_{*jn}$ is the  diamagnetic drift frequency due to $N_0(x)$,    $\alpha_i=(1-\omega_{*pi}/\omega_{dr})\simeq 1+(T_i/T_e)(1+\eta_i)$, $\omega_{*pi}=\omega_{*in}(1+\eta_i)$ and  $\eta_i=\nabla \ln T_i/\nabla\ln N_0$. 
Equations (\ref{eq:zf_fd}) and (\ref{eq:zf_mi}) are the desired result with both $\delta\phi_z$ generating mechanisms included on the same footing. Meanwhile, the equation for generating DW sidebands via coupling to $\delta\phi_Z$ is given by \cite{LChenPoP2000,PGuzdarPoP2001}
\begin{eqnarray}
\epsilon_d\phi_d=\frac{c}{B_0}\frac{k_y}{\omega_{dr}}\phi_d\frac{\partial}{\partial x}\left(\phi_{Zb}+\phi_{Zs}\right),\label{eq:dw_nl}
\end{eqnarray}
where 
\begin{eqnarray}
\epsilon_d\simeq 1-\alpha_i\rho^2_s\nabla^2_{\perp}-\frac{\omega_{*en}}{\omega_{dr}}+i\frac{\omega_{*en}}{\omega^2_{dr}}\frac{\partial}{\partial t} \label{eq:dw_dr}
\end{eqnarray}
is the DW linear dispersion relation operator with $\rho_s=c_s/\Omega_{ci}$ and $c_s=\sqrt{T_e/m_i}$. Equations (\ref{eq:zf_fd}) - (\ref{eq:dw_dr}) form the complete set of equations describing self-consistent DW-ZF nonlinear couplings in a slab plasma.     Equations (\ref{eq:zf_fd}) - (\ref{eq:dw_dr}) can be cast  in the coordinate-free form, and  extended to toroidal plasmas with effects associated with toroidal  geometry properly accounted for, as shown in the appendix.

As an application of the above theoretical results, we investigate excitation of zero-frequency zonal flow by modulational instability, which has been  considered previously without the beat-driven $\delta\phi_{Zf}$ \cite{LChenPoP2000,PGuzdarPoP2001}. Thus, we let $\phi_d=A_0 +A_+\exp(\gamma_Z t+ik_Z x) +A^*_- \exp(\gamma_Zt -ik_Z x)$ and $\phi_Z=A_Z \exp(\gamma_Zt+ik_Z x)$. Here, $A_0$, $A_+$ and $A_-$ correspond, respectively, to the finite-amplitude eDW pump mode $\Omega_0=(\omega_0,\mathbf{k}_0=k_y\hat{\mathbf{y}}+k_{\parallel}\hat{\mathbf{z}})$,  and the upper and lower sidebands  $\Omega_{\pm}=(\pm\omega_0+i\gamma_Z,\mathbf{k}_{\pm}=k_Z\hat{\mathbf{x}} \pm\mathbf{k}_0)$. Equations (\ref{eq:zf_fd}) and (\ref{eq:zf_mi}) then readily yield
\begin{eqnarray}
A_{Zb}&=&-i\frac{c k_y\omega_{*en}}{B_0k_Z\rho^2_s\omega^2_{0r}} \left(A_0A_-+A^*_0A_+\right),\label{eq:zf_am_fd}\\
\gamma_ZA_{Zs}&=&-\frac{c}{B_0} k_yk_Z \alpha_i\left(A_0A_--A^*_0A_+\right).\label{eq:zf_am_mi}
\end{eqnarray}
Meanwhile, Eq. (\ref{eq:dw_nl}) yields
\begin{eqnarray}
\epsilon_{d\pm}A_{\pm}=i\frac{ck_yk_Z}{B_0\omega_{0r}} \left(A_{Zb}+A_{Zs}\right)\left(A_0 \atop A^*_0\right),
\end{eqnarray}
where, taking $\omega_{0r}=\omega_{*en}/(1+\alpha_i b_{ys})$ with $b_{ys}\equiv k^2_y\rho^2_s$,
\begin{eqnarray}
\epsilon_{d\pm}=\alpha_i b_{Zs}\pm i\gamma_Z(\omega_{*en}/\omega^2_{0r}).  \label{eq:dw_dr_pm}
\end{eqnarray}
From Eqs. (\ref{eq:zf_am_fd}) - (\ref{eq:dw_dr_pm}), one straightforwardly derives the following modulational instability dispersion relation
\begin{eqnarray}
\left(\frac{\gamma_Z\omega_{*en}}{\omega^2_{0r}}\right)^2=\alpha_i b_{Zs}\left[\left(\alpha_b+\alpha_s\right)\left|\overline{A}_0\right|^2 -\alpha_i b_{Zs}\right], \label{eq:mi_gr}
\end{eqnarray}
where $\overline{A}_0=eA_0/T_e$,  $b_{Zs}=k^2_Z\rho^2_s$,
and 
\begin{eqnarray}
\alpha_b=\alpha_s = 2 b_{ys} (\omega_{*en}/\omega_{0r}) (\Omega_{ci}/\omega_{0r})^2.
\end{eqnarray}
Correspondingly, the instability threshold becomes 
\begin{eqnarray}
|\overline{A}_0|^2_{th} =\alpha_i b_{Zs}/(\alpha_b+\alpha_s). 
\end{eqnarray}
Noting that $\alpha_b$ and $\alpha_s$ correspond, respectively,  to beat-driven $\delta\phi_{Zb}$ and spontaneous $\delta\phi_{Zs}$ contributions and that $\alpha_b =  \alpha_s$, Eq. (\ref{eq:mi_gr}) demonstrates that $\delta\phi_{Zb}$ makes a significant  $O(1)$  contribution to the  instability drive and, thus,  the reduced   instability threshold.  Physically, this enhancement can be understood as $\delta\phi_{Zb}$ gives rise to a nonlinear frequency shift which tends to reduce  the frequency mismatch.

In summary, we have derived, using the slab plasma model, a set of equations describing nonlinear interactions between electron drift waves and zero-frequency zonal flow. Our results include, on the same footing, both the beat-driven and spontaneous excitations of zonal flow. Applying the results to the modulational instabilities, it is found that the   thresholdless   beat-driven zonal flow further  enhances significantly the   spontaneous excitation process via the enhanced modulational  instability drive.

While we adopted here, as a paradigm model, eDWs in a slab plasma to simplify the analysis and, thereby, elucidate the underlying physics mechanisms, the present results do carry important implications to realistic tokamak plasmas. It suggests that one needs to include both generating mechanisms in understanding and quantifying the excitations of zonal flows not only for drift waves; e.g., ion-temperature-gradient (ITG) and trapped-electron (TEM) modes \cite{LChenPoP2000,HChenPRL2022}, but also for shear Alfv\'en instabilities; e.g., various Aflv\'en eigenmodes and energetic-particle-modes \cite{LChenPRL2012,ZQiuNF2016,ZQiuNF2017}, where mode structures play crucial roles.  These interesting topics will be investigated in the future.   In this regard, since the zero-frequency zonal state (ZS) can be considered as a nearby nonlinear equilibrium \cite{LChenNF2007a,MFalessiNJP2023}, such studies, both analytically and numerically, may employ the recently developed comprehensive framework on the transport of phase-space zonal structures \cite{MFalessiNJP2023}.   Finally, we note that, since modulational instability is intrinsically associated with the formation and propagation of solitons, the present results also suggest that including both mechanisms could lead to the formation of drift-wave solitons at a lower drift-wave amplitude \cite{ZGuoPRL2009,NChenPoP2024}. This expectation, again,  will be a subject of a future investigation.

{\bf Acknowledgement} This work was  supported by  the National Science Foundation of China under Grant Nos. 12275236 and 12261131622, and  Italian Ministry for Foreign Affairs and International Cooperation Project under Grant  No. CN23GR02.
 This work was also supported by the EUROfusion Consortium, funded by the European Union via the Euratom Research and Training Programme (Grant Agreement No. 101052200 EUROfusion). The views and opinions expressed are, however, those of the author(s) only and do not necessarily reflect those of the European Union or the European Commission. Neither the European Union nor the European Commission can be held responsible for them.

\section*{Author declarations}
\section*{Conflict of interest}  
The authors have no conflicts to disclose.
\section*{Date availability statement} 
The data that support the findings of this study are available from the corresponding author upon reasonable request.

\appendix
\section{Coordinate-free form of the coupled DW-ZF equations}\label{sec:app}

Denoting $\Phi_k=e\delta\phi_k/T_e$, the coupled DW-ZF equations, Eqs. (\ref{eq:zf_fd}) - (\ref{eq:dw_dr}),  can be cast  in the following coordinate-free form
\begin{widetext}
\begin{eqnarray}
\nabla^2\Phi_{zb}&=&-\left(\frac{c_s}{\omega_{0r}}\right)^2\left\{\nabla\Phi_d\times\mathbf{b}\cdot\nabla\left[\nabla\Phi^*_d\times \mathbf{b}\cdot\nabla\ln N_0\right] + c.c.\right\},\label{eq:A1}\\
\frac{\partial}{\partial t}\nabla^2\Phi_{zs}&=& \frac{c^2_s}{\Omega_{ci}}\alpha_i \nabla\cdot \left[\left( \nabla\Phi_d\times \mathbf{b}\cdot\nabla\right)\nabla\Phi^*_d +c.c.\right],\label{eq:A2}\\
\overline{\epsilon}_d\Phi_d&=& - i \frac{c^2_s}{\Omega_{ci}\omega_{dr}}\left(\nabla\Phi_d\times\mathbf{b}\right)\cdot\nabla \left(\Phi_{zb}+\Phi_{zs}\right),\label{eq:A3}
\end{eqnarray}
and 
\begin{eqnarray}
\overline{\epsilon}_d=1-\alpha_i \rho^2_s\nabla^2_{\perp} - i \left(\frac{c^2_s}{\Omega_{ci}\omega_{dr}}\right) \left(1-i\frac{\partial/\partial t}{\omega_{dr}}\right) \left(\mathbf{b}\times\nabla\ln N_0\right)\cdot\nabla. \label{eq:A4}
\end{eqnarray}
\end{widetext}

For a circular tokamak, the left hand sides of Eqs. (\ref{eq:A1}) and (\ref{eq:A2}) should be multiplied by a neoclassical  factor $1.6 q^2/\sqrt{\epsilon}$ \cite{MRosenbluthPRL1998}, with $q$ being the safety factor and $\epsilon=r/R$ being the ratio between minor and major radii.

\end{document}